\def\cm2{cm$^{-2}$}

\newcommand{\msun}{M_\odot}
\newcommand{\lsun}{L_\odot}

\documentclass[article]{has40}
\usepackage{graphicx}
\usepackage{amssymb}
\tabletypesize{\normalsize}  % AMcW

\def\plotone#1{\centering \leavevmode
\includegraphics[width=.95\columnwidth]{#1}}

\def\plotone#1{\centering \leavevmode
\includegraphics[width=.95\columnwidth]{#1}}

\shortauthors{Kuehn}
\shorttitle{RR Lyrae in the LMC}

\begin{document}
\large    %AMcW  The conference proceedings will employ large size print
\pagenumbering{arabic}
\setcounter{page}{123}

\title{RR Lyrae in the LMC: Insights Into the Oosterhoff Phenomenon}

%
% Here is an example of how to include the author names and affiliations
%
\author{{\noindent Charles Kuehn{$^{\rm 1}$}, Horace A. Smith{$^{\rm 2}$},  M\'arcio Catelan{$^{\rm 3,4}$}, Young-Beom Jeon{$^{\rm 5}$}, James M. Nemec{$^{\rm 6}$}, Alistair R. Walker{$^{\rm 7}$}, Andrea Kunder{$^{\rm 7}$}, Kyra Dame{$^{\rm 2}$}, Barton J. Pritzl{$^{\rm 8}$}, Nathan De Lee{$^{\rm 9}$}, Jura Borissova{$^{\rm 10}$}\\
\\
{\it (1) Sydney Institute for Astronomy, University of Sydney, Sydney, Australia\\(2) Department of Physics and Astronomy, Michigan State University, East Lansing, MI, USA\\(3) Pontificia Universidad Cat$\rm{\acute{o}}$lica de Chile, Facultad de F\'{i}sica, Departamento de Astronom\'ia y Astrof\'isica, Santiago, Chile\\(4) The Milky Way Millennium Nucleus, Santiago, Chile\\(5) Korea Astronomy and Space Science Institute, Daejeon, Korea\\(6) Department of Physics and Astronomy, Camosun College, Victoria, British Columbia, Canada\\(7) Cerro Tololo Inter-American Observatory, National Optical Astronomy Observatory, La Serena, Chile\\(8) Department of Physics and Astronomy, University of Wisconsin Oskosh, Oshkosh, WI, USA\\(9) Department of Physics and Astronomy, Vanderbilt University, Nashville, TN, USA\\(10) Departamento de F\'isica y Astronom\'ia, Falcultad de Ciencias, Universidad de Valpara\'iso, Valpara\'iso 
}
}}

%
% And here is how to add the e-mail addresses
%
\email{(1) kuehn@physics.usyd.edu.au (2) smith@pa.msu.edu, damekyra@msu.edu (3) mcatelan@astro.puc.cl (5) ybjeon@kasi.re.kr (6) nemec@camosun.bc.ca (7) awalker@ctio.noao.edu, akunder@ctio.noao.edu (8) pritzlb@uwosh.edu (9) nathan.delee@vanderbilt.edu (10) jura.borissova@uv.cl}

% If you really you need to add an alternate institution, then update and uncomment the following line.
% It's not very pretty though
%\altaffiltext{}{(3) also at The Institute for the Insane}

\begin{abstract}
Although more than eight decades have passed since P. Th. Oosterhoff drew attention to differences in the properties of RR Lyrae variables in globular clusters, the origin and significance of the Oosterhoff groups remain unclear.  Nonetheless, the accumulation of extensive new observations of RR Lyrae stars in globular clusters of the Milky Way and Local Group galaxies allows a fresh look at the phenomenon.  Insights come not only from surveys of variables within the original Oosterhoff groups I and II but also from recent observations of the Oosterhoff-intermediate systems found especially in smaller Local Group galaxies.  We will compare properties of RR Lyrae in several systems to investigate what they reveal about system-to-system differences of transition temperature between fundamental-mode and first overtone pulsators and of horizontal branch luminosity.  Both transition temperature and horizontal branch luminosity have at various times been credited as playing roles in the creation of the Oosterhoff dichotomy.
\end{abstract}

\section{Introduction}
\citet{oosterhoff39} noticed that globular clusters in the Milky Way could be divided into two groups based on the average periods and number fraction of their RR Lyrae stars.  Although his original sample size was only five globular clusters, additional analysis \citep{oosterhoff44, sawyer44} confirmed that this division existed, and these groups were subsequently named after Oosterhoff.  Oosterhoff I (Oo-I) globular clusters tend to have shorter average periods for their RR Lyraes, and are more metal-rich than Oosterhoff II (Oo-II) clusters \citep{arp55,smith95}.  The number fraction (the ratio of number of first overtone dominant RR Lyrae to total RR Lyrae) tends to be lower in Oo-I clusters.

\begin{figure*}
\centering
\plotone{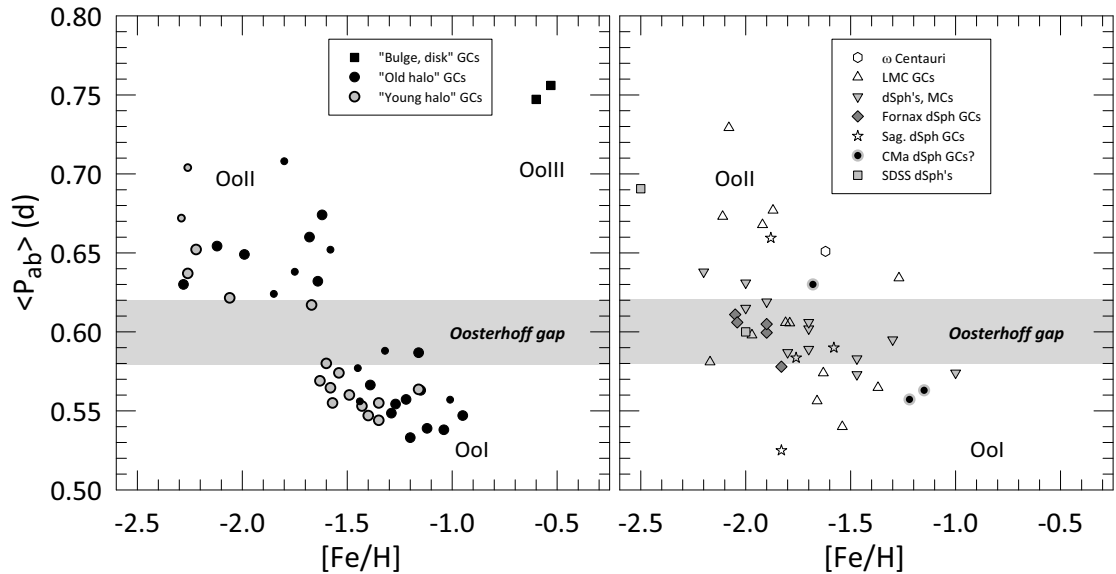}
\vskip0pt
\caption{Average period for RRab stars $\langle P_{ab}\rangle$ vs [Fe/H] for globular clusters in the Milky Way (\textit{left}) and nearby dwarf galaxies and their globular clusters ($right$).  Figure reprinted from \citet{catelan09b} with permission.}
\label{oostfig}
\end{figure*}

A plot of the average RRab period ($\langle P_{ab}\rangle$) vs cluster metallicity for Milky Way globular clusters (Figure \ref{oostfig}, left panel) clearly shows the two Oosterhoff groups as well as a zone of avoidance between them, called the ``Oosterhoff gap''.  The right panel of Figure \ref{oostfig} shows $\langle P_{ab}\rangle$ vs [Fe/H] for stellar systems in Local Group dwarf galaxies.  When one includes Local Group dwarf galaxies, the Oosterhoff gap disappears as stellar systems populate this region, as seen in Figure \ref{oostfig}.   In fact, these extragalactic objects seem to preferentially lie in the gap \citep{catelan09b}.  

These Oosterhoff-intermediate (Oo-Int) objects, as the globular clusters and/or dwarf galaxies that fall into the gap are referred, present a challenge to the accretion model for the formation of the Milky Way halo \citep{bullockjohnston05,abadi06,font06}.  Of the $43$ Milky Way globular clusters that contain at least five identified RR Lyrae stars, only four of these clusters can be classified as Oo-Int, and they are on the edges of the Oosterhoff gap.  Looking at the nearby dwarf galaxies and their globular clusters, of the $36$ objects with at least five identified RR Lyrae stars, $17$ of them are classified as Oo-Int.  Performing a Kolmogorov-Smirnov test reveals that there is only a $1.8\%$ chance that these two sets of objects are from the same parent population \citep{catelan09a}.  If the Milky Way halo formed by accreting objects like the early counterparts of the current dwarf galaxies, then we would expect to see RR Lyrae stars in the halo with the same Oosterhoff classification as we see in the dwarfs.  The presence of the Oosterhoff gap when looking at the globular clusters in the Milky Way halo indicates that if the halo formed from accreting dwarf galaxies, these accreted galaxies did not look like our current dwarfs \citep{catelan09a}.  The Oosterhoff dichotomy is also present in the field stars of the Milky Way halo which consists primarily of an Oo-I population with a significantly smaller Oo-II population \citep{vivas04,kinemuchi06,drake13}.

Clearly one must understand the Oosterhoff phenomenon if one is to fully understand Milky Way formation.  Our proximity to Milky Way globular clusters compared to extragalactic sources means that in general these objects are better surveyed.  This results in Oo-I and Oo-II objects being more completely studied than Oo-Int ones.  To better understand the nature of the Oosterhoff phenomenon, we targeted a series of globular clusters in the Large Magellanic Cloud to obtain well-sampled, high-photometric precision light curves for the RR Lyrae stars.  The details of these observations and the results for the individual target clusters can be found in NGC 1466:\citet{kuehn11}, NGC 1786:\citet{kuehn12}, Reticulum:\citet{kuehn13a}, and NGC 2210:\citet{jeon13}.

\section{Bailey Diagram Behavior of Oosterhoff-Intermediate Objects}

In a Bailey (period-amplitude) diagram, RR Lyrae stars in Oo-I and Oo-II fall into distinct, separate positions.  In general, Oo-I RR Lyrae stars are still on the zero-age horizontal branch (ZAHB), but some stars may have evolved off the ZAHB and appear in the same location on the Bailey diagram as Oo-II stars \citep{cacciari2005}.  The pulsational characteristics of the RR Lyrae stars allows for the Oosterhoff trend loci to be determined for RRab and RRc stars in Oo-I and Oo-II clusters.  These loci can be used to identify the Oosterhoff class of clusters and even individual stars.  Do RR Lyrae stars in Oo-Int clusters also occupy similar positions on the Bailey diagram and can a locus for them be determined?  To answer this we compare the Bailey diagram behavior of several different Oo-Int systems, the properties of which are summarized in Table \ref{gccompare}.

%RR Lyrae stars in Oo-II clusters tend to occupy similar positions on Bailey (period-amplitude) diagrams.  In Oo-I clusters RR Lyrae stars which are still on the zero age horizontal branch (ZAHB) do likewise, though at a different position from the Oo-II stars; RR Lyrae stars Oo-I clusters which have evolved off of the ZAHB occupy positions more similar what is seen in Oo-II clusters \citep{cacciari2005}.  The similarity in behavior of the RR Lyrae stars allows for loci to be determined for RRab and RRc stars in Oo-I and Oo-II clusters; these loci can be used to determine the Oosterhoff class of other clusters and even individual stars.  Do RR Lyrae stars in Oo-Int clusters also occupy similar positions on the Bailey diagram and can a locus for them be determined?  To answer this we compare the Bailey diagram behavior of several different Oo-Int systems, the properties of which are summarized in Table \ref{gccompare}.

\begin{flushleft}
\begin{deluxetable*}{lccccccc}
\tabletypesize{\normalsize}
\tablecaption{Properties of Stellar Systems Examined in Section 2}
\tablewidth{0pt}
\tablehead{\\ \colhead{Object} & \colhead{[Fe/H]} & \colhead{HB Type} & \colhead{$\langle P_{ab}\rangle$} & \colhead{$\langle P_{c}\rangle$} & \colhead{$P_{ab,min}$}& \colhead{$P_{c,max}$} & \colhead{$f_{cd}$} \\
}
\startdata
NGC 1466 & -1.64 & 0.38 & 0.591 & 0.335 & 0.4934 & 0.3770 & 0.39\\
NGC 2210 & -1.65 & 0.65 & 0.612 & 0.350 & 0.5105 & 0.4355 & 0.35\\
Draco & -2.1 & & 0.615 & 0.375 & 0.5366 & 0.4310 & 0.21\\
%NGC 2257 & -1.95 & 0.46 & 0.611 & 0.333 & 0.5042 & 0.3823 & 0.50\\
M3 & -1.55 & 0.18 & 0.555 & 0.340 & 0.4560 & 0.4860 & 0.21\\
M15 & -2.26 & 0.67 & 0.644 & 0.370 & 0.5527 & 0.4406 & 0.58\\
\enddata
\label{gccompare}
\end{deluxetable*}
\end{flushleft}

\begin{figure*}
\centering
\epsscale{0.7}
\plotone{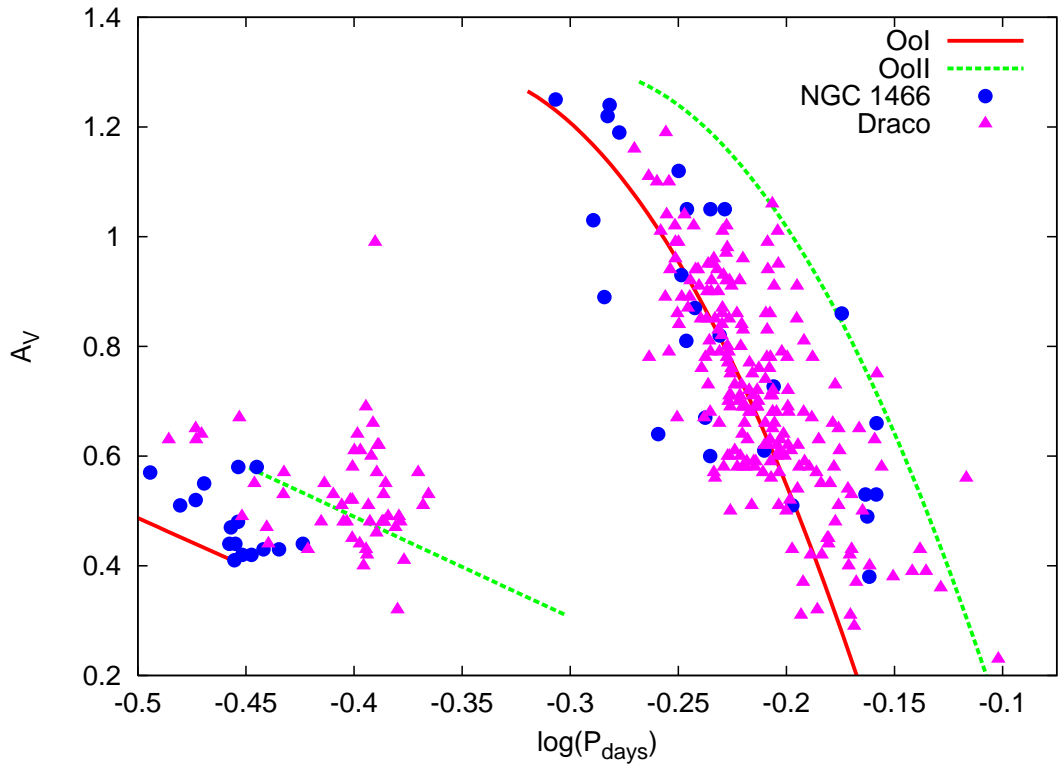}
\vskip0pt
\caption{Bailey diagram, log period vs $V$-band amplitude, for the RR Lyrae stars in NGC 1466 (blue circles) and Draco (purple triangles).  Red and green lines indicate the typical position for RR Lyrae stars in Oosterhoff I and Oosterhoff II clusters, respectively \citep{cacciari2005,zorotovic10}.  Data for NGC 1466 are from \citet{kuehn11} while data for Draco are from \citet{kinemuchi08}.}
\label{1466dracoperamp}
\end{figure*}

We first compare the Bailey diagram of NGC 1466 \citep{kuehn11}, a globular cluster in the LMC, and the Draco dwarf spheroidal galaxy \citep{kinemuchi08}.  Both objects are Oo-Int with Draco being more metal-poor ([Fe/H]$=-2.1$) than NGC 1466 ([Fe/H]$=-1.64$).  In Figure \ref{1466dracoperamp} we see that the RRab stars in Draco and NGC 1466 occupy similar positions on the diagram and show a similar level of scatter.  The main difference in the RRab stars is that in NGC 1466 the RRab's extend down to shorter periods ($P_{ab,min,1466}=0.4934$ days) than they do in Draco ($P_{ab,min,Draco}=0.5366$ days).  Since Draco contains more RR Lyrae stars than NGC 1466, Draco contain 214 RRab's while NGC 1466 contains 30, this is unlikely to be due to statistical effects.  In contrast to the RRab stars, the RRc stars in these two objects occupy very different positions on the Bailey diagram, with the RRc in NGC 1466 having shorter periods than those in Draco.

\begin{figure*}
\centering
\plotone{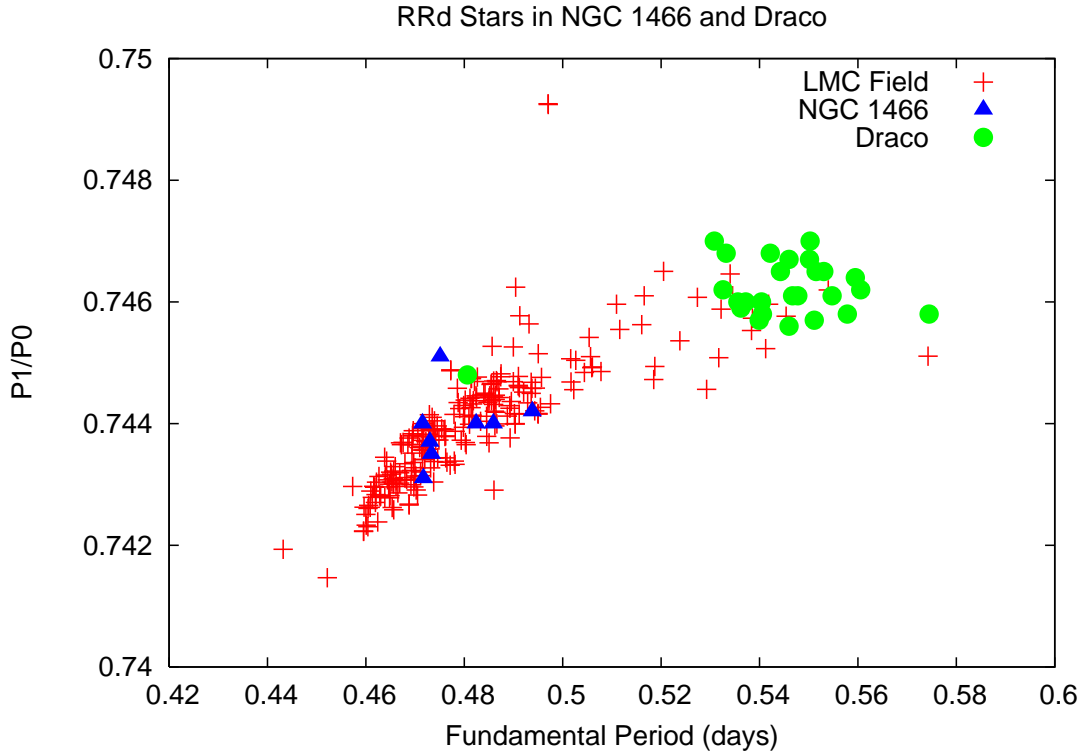}
\vskip0pt
\caption{Petersen diagram showing the ratio of the first overtone period to the fundamental-mode period vs fundamental-mode period for the RRd stars in NGC 1466 (blue triangles) and Draco (green circles).  Also plotted are the RRd stars in the LMC field (red plus symbols) from \citet{soszy03}.}
\label{1466dracopetersen}
\end{figure*}

The RRd stars in both objects also show different characteristics.  Figure \ref{1466dracopetersen} shows the RRd stars for both NGC 1466 and Draco plotted on a Petersen diagram.  All but one of the RRd stars in Draco have longer fundamental-mode periods and higher period ratios than those in NGC 1466.  The RRd stars in Draco occupy a similar position to what is typically seen in Oo-II clusters while those in NGC 1466 enjoy a position similar to those in Oo-I objects \citep{popielski00}.

\begin{figure*}
\centering
\plotone{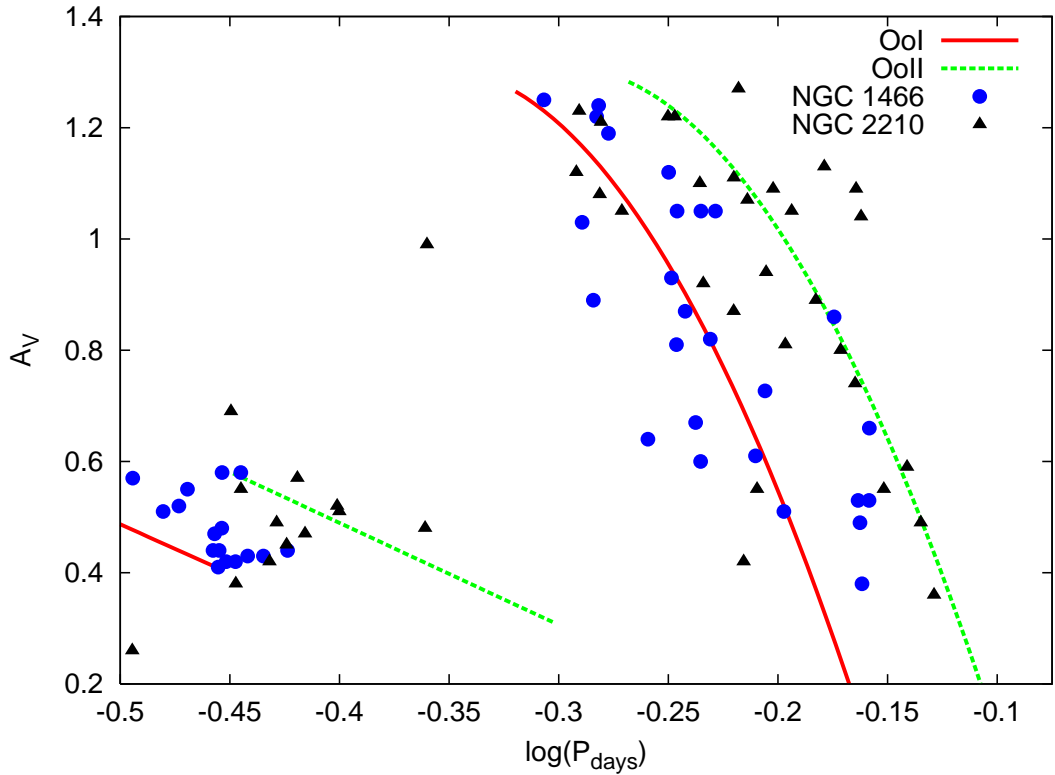}
\vskip0pt
\caption{Bailey diagram, log period vs $V$-band amplitude, for the RR Lyrae stars in NGC 1466 (blue circles) and NGC 2210 (black triangles).  Data for NGC 2210 is from \citet{jeon13}.}
\label{1466vs2210peramp}
\end{figure*}

The difference in metallicity between NGC 1466 and Draco cannot be the sole explanation for the differing Bailey diagram behavior of Oo-Int objects.  This is immediately apparent when one compares the Bailey diagrams for NGC 1466 and NGC 2210 (Figure \ref{1466vs2210peramp}), which are both Oo-Int clusters in the LMC with metallicities that are nearly identical (${\rm [Fe/H]}\approx-1.65$).  Unlike NGC 1466 and Draco, both of which had RRab stars that clumped more toward the Oo-I locus, the RRab stars in NGC 2210 clump more towards the Oo-II line \citet{jeon13}. Although there is a great deal of scatter, a similar trend is seen in the RRc stars.  The locations of the RR Lyrae stars between these two clusters looks very similar to the difference in position between an Oo-I cluster and an Oo-II cluster.

%\begin{figure*}
%\centering
%\plotone{2210_2257_vperamp.ps}
%\vskip0pt
%\caption{Bailey diagram, log period vs $V$-band amplitude, for the RR Lyrae stars in NGC 2210 (black triangles) and NGC 2257 (blue circles).  Data for NGC 2257 are from \citet{nemec09}.}
%\label{2210vs2257peramp}
%\end{figure*}

%We also look at the Bailey diagram behavior of NGC 2257, another Oo-Int globular cluster in the LMC which has a metallicity of ${\rm [Fe/H]}=-1.95\pm0.02$ dex \citep{mucciarelli10}.  Figure \ref{2210vs2257peramp} shows that the RRab stars in NGC 2257 display behavior similar to those in NGC 2210, clustering toward the Oo-II line but with a great deal of scatter.  Despite the similarity in the behavior of the RRab stars, the RRc stars show a difference in position, with the RRc in NGC 2257 being shifted to shorter periods than their counterparts in NGC 2210.

\subsection{Oosterhoff-Intermediate vs Oosterhoff I/II Systems}

\begin{figure*}
\centering
\plotone{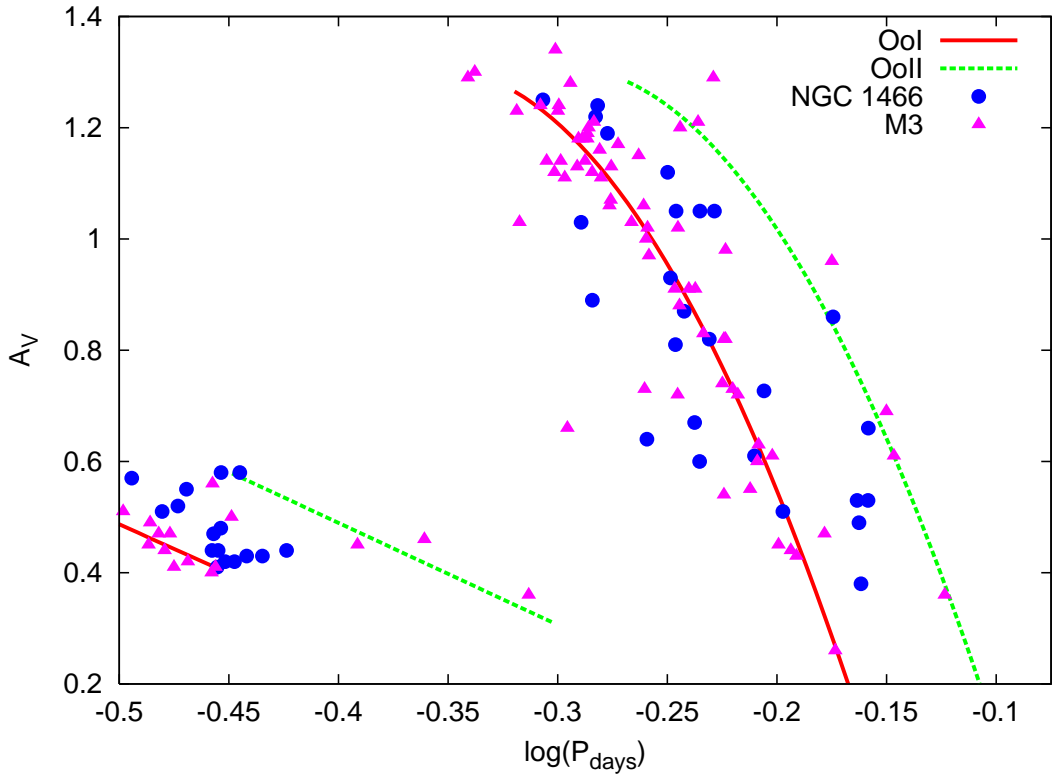}
\vskip0pt
\caption{Bailey diagram, log period vs $V$-band amplitude, for the RR Lyrae stars in NGC 1466 (blue circles) and M3 (purple triangles).  Data for M3 come from \citet{cacciari2005}.}
\label{1466vsm3peramp}
\end{figure*}

Since NGC 1466 and NGC 2210 feature RRab stars that tend toward either the Oo-I or Oo-II loci, respectively, in the Bailey diagram, we examine how these two clusters compare to bona fide Oo-I/II clusters.  Figure \ref{1466vsm3peramp} shows the combined Bailey diagram for the RR Lyrae stars in NGC 1466 and M3 \citep{cacciari2005}, an Oo-I cluster located in the halo of the Milky Way.  M3 has a metallicity of ${\rm [Fe/H]}=-1.55\pm0.13$ \citep{smolinski11}, somewhat more metal-rich than NGC 1466. The RRab stars in the two clusters occupy similar positions, though the majority of the RRab stars in M3 show tighter clustering around the Oo-I locus than those in NGC 1466.  The other difference in the RRab stars is that the RRab in M3 extend to shorter periods than those in NGC 1466; the minimum period of an RRab star in M3 is $0.4560$ days while in NGC 1466 it is $0.4934$ days.  More notably, a majority of M3 RRc stars have a smaller scatter about the Oo-I locus and are shifted toward a shorter period than their counterparts in NGC 1466.  This shift in RRc periods, along with M3 having a shorter $P_{ab,min}$, supports the transition between RRab and RRc stars in M3 occurring at a shorter period than in NGC 1466.

\begin{figure*}
\centering
\plotone{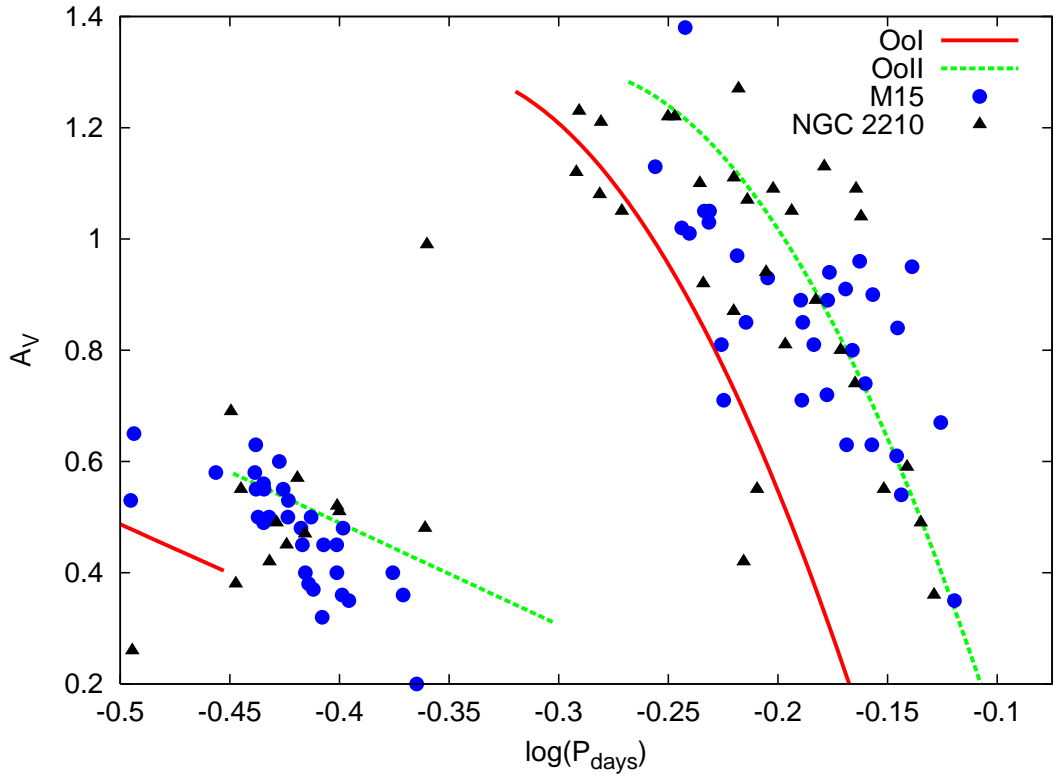}
\vskip0pt
\caption{Bailey diagram, log period vs $V$-band amplitude, for the RR Lyrae stars in NGC 2210 (black triangles) and M15 (blue circles).  Data for M15 is from \citet{corwin08}.}
\label{2210vsm15peramp}
\end{figure*}

We also compare NGC 2210 to M15 \citep{corwin08}, a Milky Way Oo-II cluster with a metallicity of ${\rm [Fe/H]}=-2.26$ \citep{harriscat}.  Figure \ref{2210vsm15peramp} shows the Bailey diagram for these two clusters.  The RRab stars in both clusters show a great deal of scatter and largely occupy similar positions.  The major difference in the RRab stars between the two clusters is that the RRab stars in NGC 2210 extend to a shorter period than those in M15; $P_{ab,min}=0.5105$ days in NGC 2210 and $0.5527$ days in M15.  If one removed the five NGC 2210 RRab stars that have a $\log(P)\le -0.25$, the distributions of RRab stars in the two clusters would be essentially the same.  The shorter-period RRab to RRc transition in NGC 2210 seems to be responsible for the Oo-Int classification of the cluster.

When compared to bona fide Oo-I/II clusters, the RRab stars in NGC 1466 and NGC 2210 seem to occupy similar positions in the Bailey diagram to the RRab stars in the comparison cluster.  While the Oo-Int clusters show slightly more scatter in the position of their RRab stars, the factor that seems to drive the Oosterhoff classification of these clusters is the transition period between RRab and RRc stars, as indicated by the minimum period of the RRab stars.

\section{RR Lyrae Physical Properties}

\citet{jurcsikkovacs96, jurcsik98, simonclement93} have shown that the Fourier parameters of RR Lyrae light curves can be used to estimate their physical properties.  The RRab light curves were fit with a Fourier series of the form
\begin{equation}
m(t)=A_{0}+\sum_{j=1}^{n}A_{j}\sin (j\omega t+\phi_{j}),
\end{equation}
while the RRc light curves were fit with a cosine series of similar form.  The resulting Fourier coefficients and the relations from \citet{jurcsikkovacs96}, \citet{jurcsik98}, Kov\'acs \& Walker (1999, 2001), \citet{simonclement93}, and \citet{morgan07} were used to calculate the physical properties of the RR Lyrae stars.  As a note of caution, the \citet{simonclement93} relations for RRc stars produce values that are in violation of the period-mean density relation \citep{catelan04,debsingh10}; despite this, the resulting values can still be used for comparisons between clusters.  For further details on this method and the results from the individual clusters in our study, please see Kuehn at al. (2011, 2012, 2013) and \citet{jeon13}.

\begin{figure*}
\centering
\epsscale{0.7}
\plotone{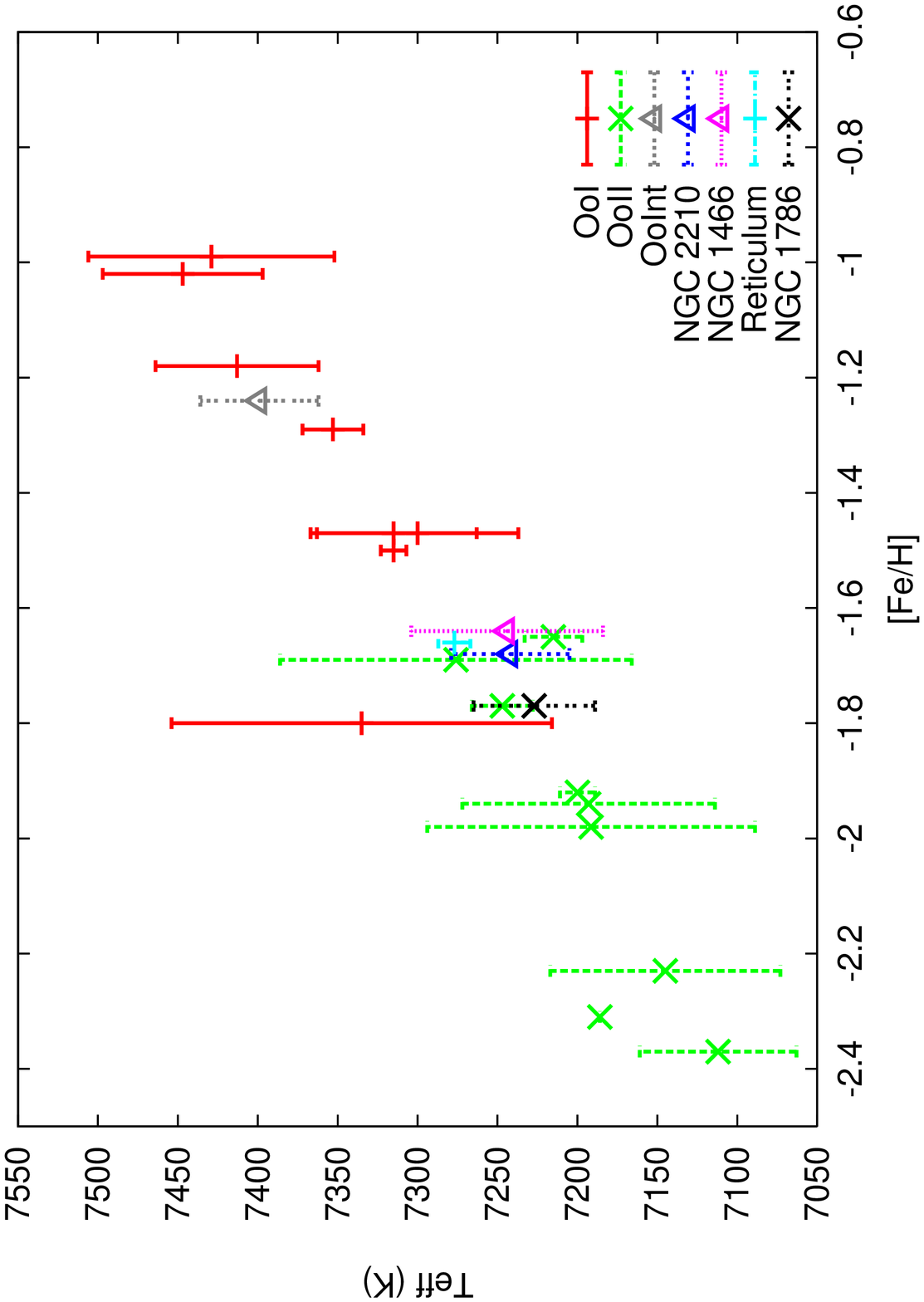}
\vskip0pt
\caption{The average effective temperature, $\langle T_{\rm eff}\rangle$, for RRc stars vs cluster metallicity.}
\label{aftercteff}
\end{figure*}

\begin{figure*}
\centering
\plotone{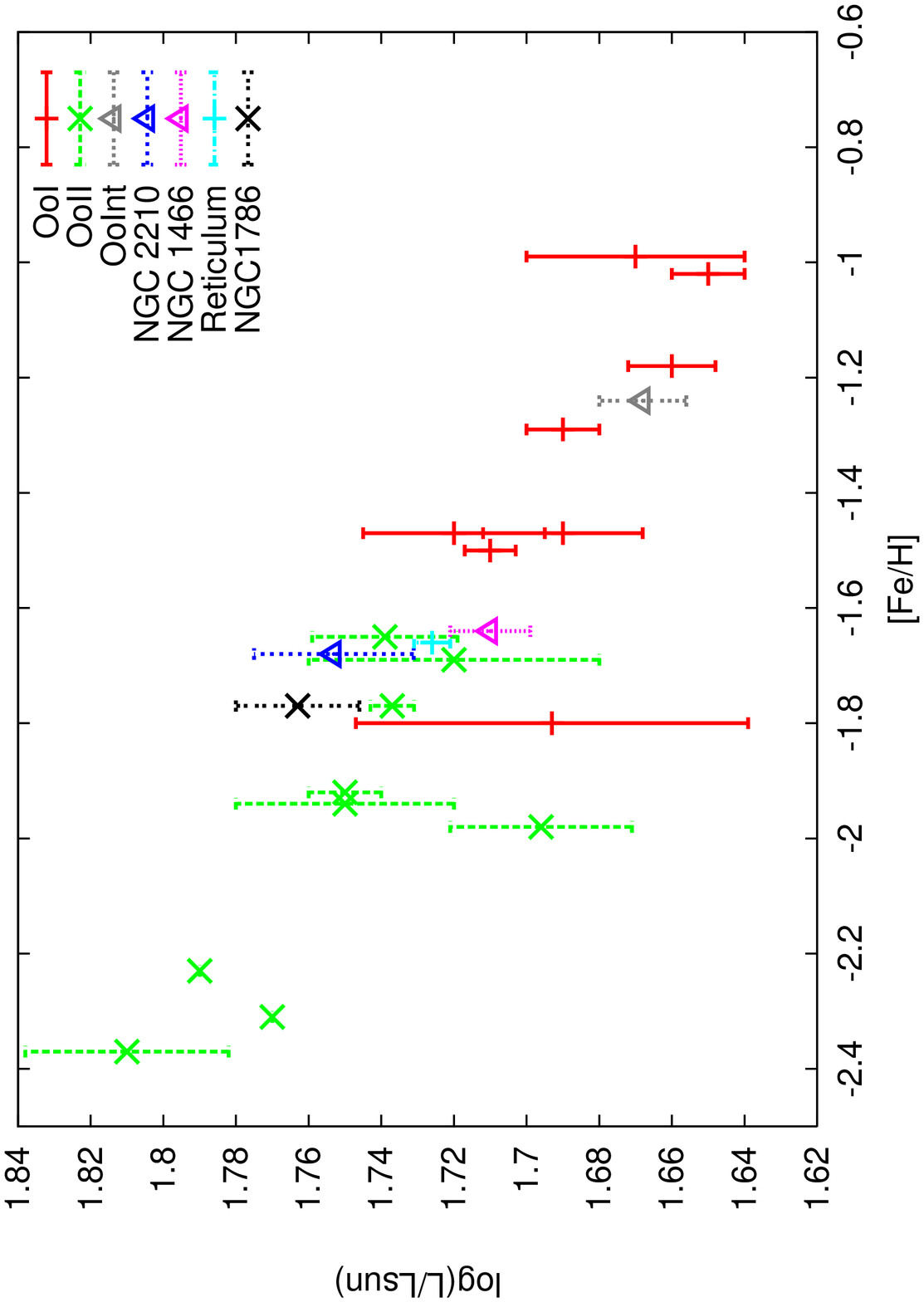}
\vskip0pt
\caption{Average luminosity (in terms of the luminosity of the Sun), $\langle L/\lsun \rangle$, for RRc stars vs cluster metallicity.}
\label{afterclum}
\end{figure*}

\begin{figure*}
\centering
\plotone{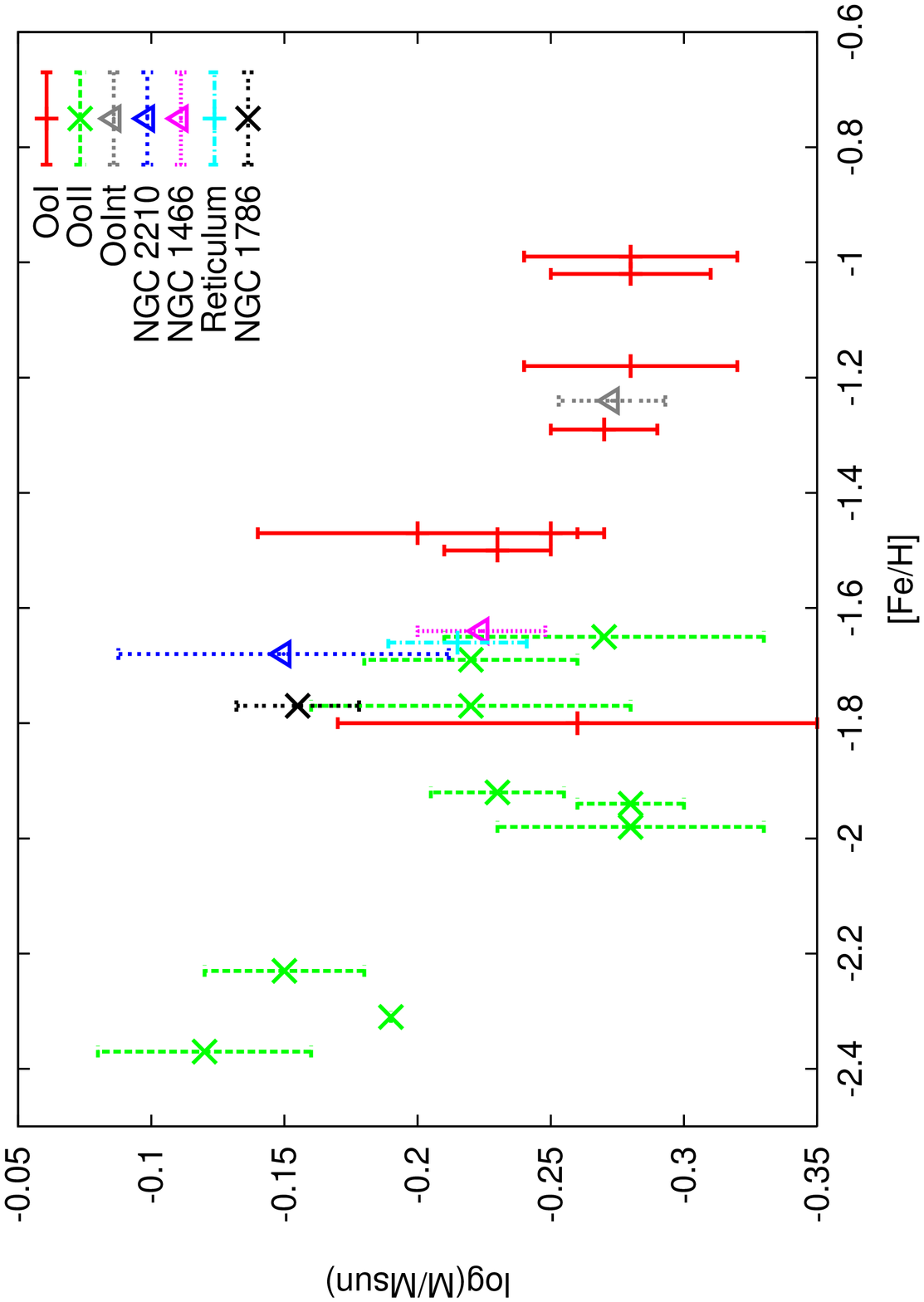}
\vskip0pt
\caption{Average mass (in terms of the mass of the Sun), $\langle M/\msun \rangle$, for RRc stars vs cluster metallicity.}
\label{aftercmass}
\end{figure*}

\begin{figure*}
\centering
\plotone{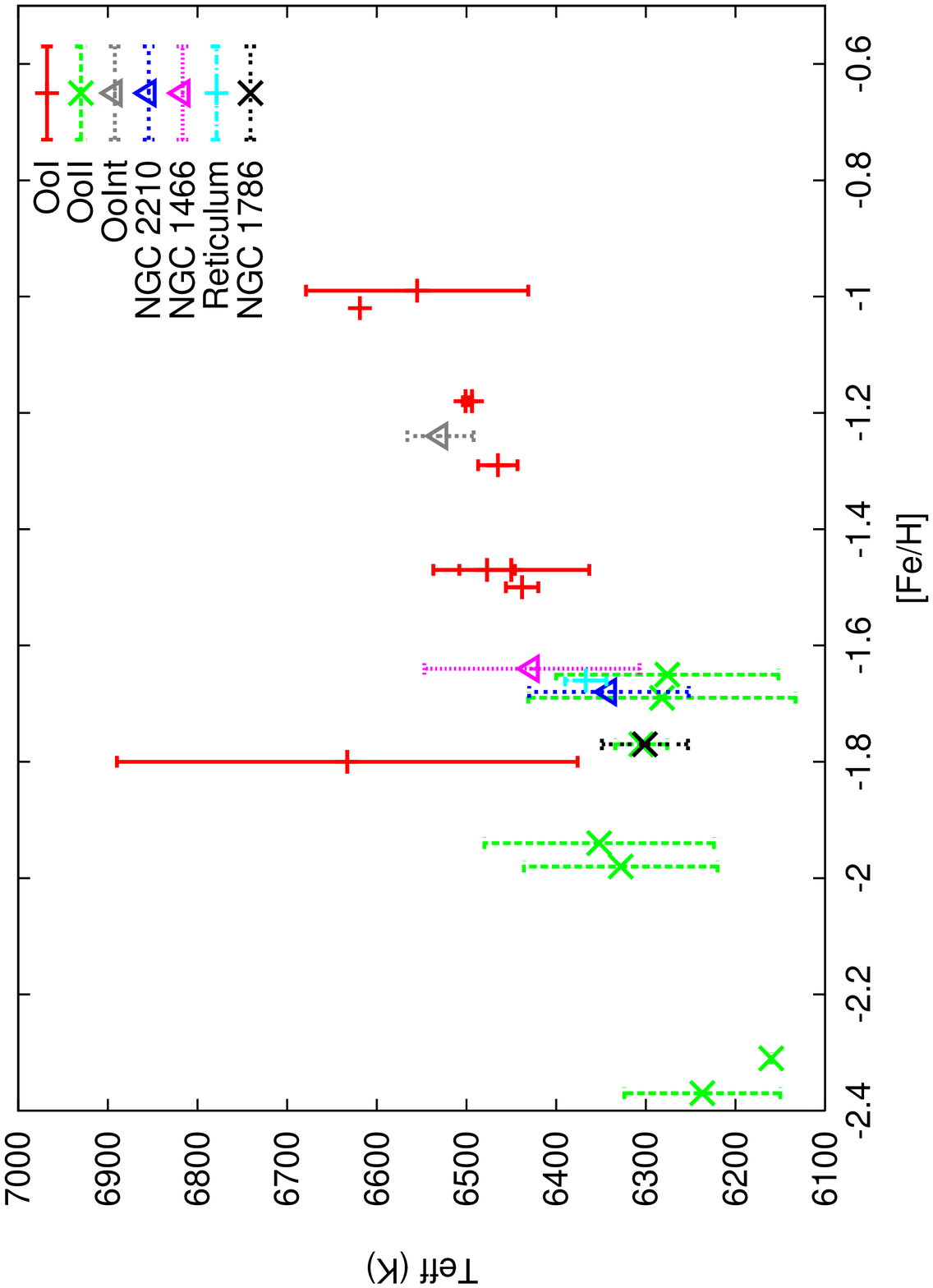}
\vskip0pt
\caption{The average effective temperature, $\langle T_{\rm eff}\rangle$, for RRab stars vs cluster metallicity.}
\label{afterabteff}
\end{figure*}

\begin{figure*}
\centering
\plotone{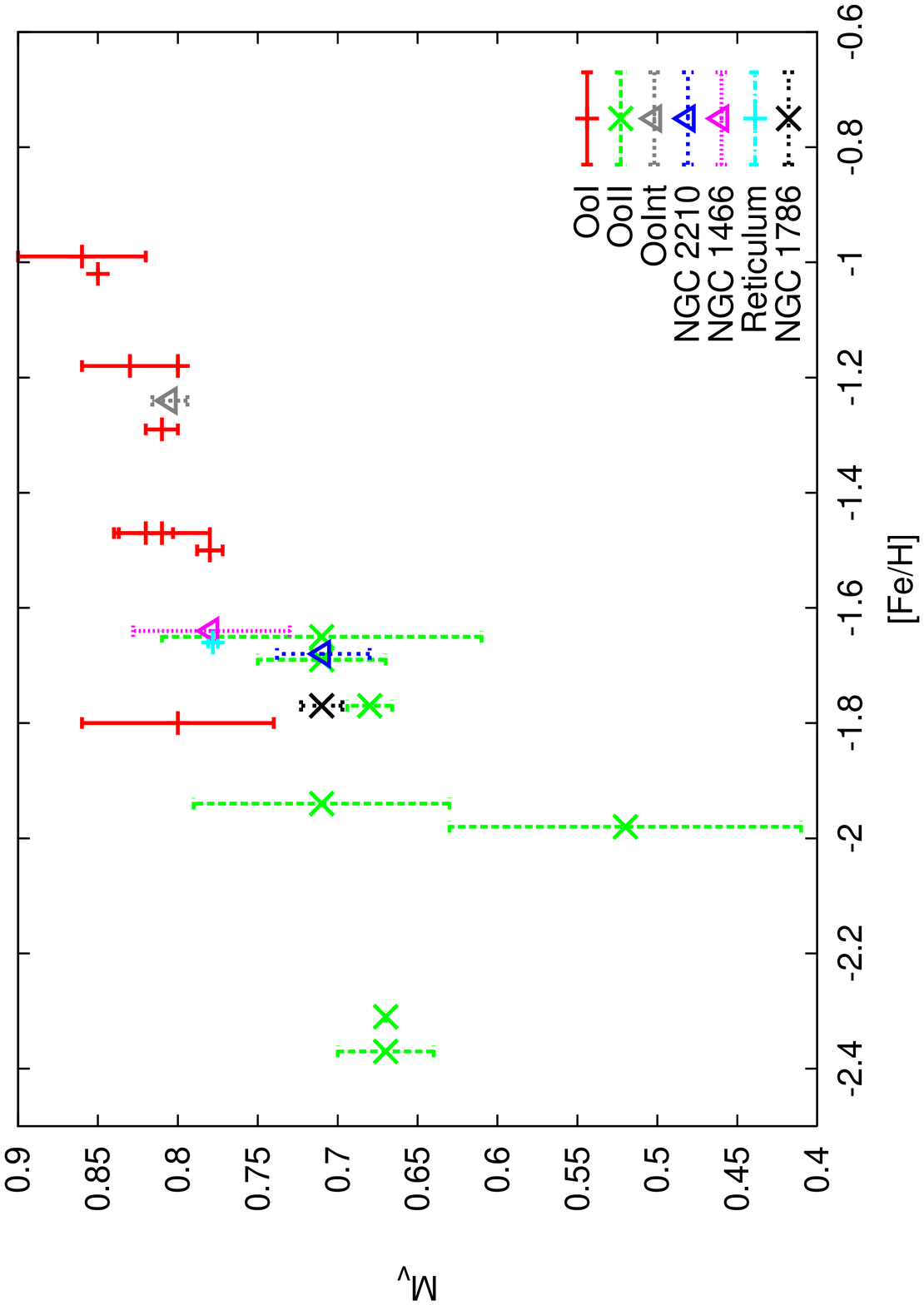}
\vskip0pt
\caption{The average $V$-band absolute magnitude, $\langle M_{V}\rangle$, for RRab stars vs cluster metallicity.}
\label{afterabmag}
\end{figure*}

Physical properties for several Milky Way globular clusters have been previously determined using this method (Contreras et al. 2010, Corwin et al. 2003, L\'{a}zaro et al. 2006, and references therein).  We  compare the physical properties for our target clusters (effective temperature, luminosity and mass of RRc stars and effective temperature and absolute magnitude of RRab stars) to these previously studied clusters to see if there are any correlations with Oosterhoff type.  Figures \ref{aftercteff}-\ref{afterabmag} show the average physical properties for the RR Lyrae stars plotted against cluster metallicity for our target clusters and the previously studied Milky Way clusters; cluster metallicities are from \citet{harriscat}.

Figures \ref{aftercteff}-\ref{afterabteff} show a roughly linear trend in the physical properties of the RR Lyrae stars versus cluster metallicity.  Globular clusters of all three Oosterhoff types appear to fall along the same linear trend in these figures, suggesting that the physical properties responsible for the formation of RR Lyrae stars are the same in Oo-I, Oo-II, and Oo-Int objects.

%The relationship between the $V$-band absolute magnitude and cluster metallicity, Figure \ref{afterabmag}, is the one exception to the linear trend in physical properties.  The RRab absolute magnitude trend appears to be more of a step function with Oo-I clusters being less luminous than Oo-II clusters; this step function relationship between metallicity and RR Lyrae luminosity was first seen in Omega Cen \citep{lee91}.  The Oo-Int clusters straddle the break with NGC 1466 and M75 (grey triangle) falling with the Oo-I clusters while NGC 2210 fits in with the Oo-II clusters.

The relationship between the $V$-band absolute magnitude and cluster metallicity, Figure \ref{afterabmag}, is the one exception to the linear trend in physical properties.  Oo-I clusters form a linear trend with slightly increasing luminosity as cluster metallicity decreases; Oo-II clusters are systematically more luminous than the Oo-I clusters and appear to have a flatter luminosity vs cluster metallicity relationship.  This step function relationship between metallicity and RR Lyrae luminosity was first seen in Omega Cen \citep{lee91}.  The Oo-Int clusters straddle the break with NGC 1466 and M75 (grey triangle) \citep{corwin03} falling with the Oo-I clusters while NGC 2210 fits in with the Oo-II clusters.

\section{Conclusion}

Unlike Oo-I or Oo-II objects, Oo-Int displays a wide variety of Bailey diagram behavior.  Some Oo-Int objects have RRab stars that are similar to those in Oo-I objects while others have RRab stars similar to Oo-II objects.  The RRc and RRd stars in Oo-Int objects also vary between being more Oo-I- or Oo-II-like but the behavior of the first overtone dominant pulsators (RRc and RRd stars) does not correlate with that of the fundamental-mode pulsators (RRabs).  The difference in the luminosity of the RR Lyrae stars between various Oo-Int objects likely plays a role in these differences.

We also looked at how the Bailey diagram behavior of Oo-Int objects compares to that of Oo-I/II systems.  We found that some Oo-Int objects have RRab stars that appear similar to those in Oo-I/II systems with the major difference being the minimum RRab period of the clusters.  This minimum RRab period, which represents the transition point between RRab and RRc stars, appears to be the important difference between whether a system is classified as an Oo-Int or an Oo-I/II.

The Fourier-derived physical properties for the RR Lyrae stars mostly show linear trends versus cluster metallicity.  These trends are the same for Oo-I, Oo-II, and Oo-Int clusters, suggesting that the physical processes that formed the RR Lyrae stars are the same regardless of Oosterhoff type.  The one exception to the linear trends that carry through all three Oosterhoff groups is the relationship between the absolute magnitude of RRab stars and cluster metallicity (Figure \ref{afterabmag}), which appears as more of a step function with RRab stars in Oo-II clusters more luminous than those in Oo-I clusters.  The Oo-Int clusters fall on either side of the step function with some falling in with the Oo-I clusters while others fall in with the Oo-II ones.  

The step-function like behavior in the absolute magnitude of the RRab stars and the differing Bailey diagram behavior of Oo-Int systems indicate that Oo-Int objects cannot be considered a homogenous set.  The difference between the Oo-I-like Oo-Int clusters and the Oo-II-like ones may be due to the luminosity difference of the RRab stars seen in Figure \ref{afterabmag}.

\section{Acknowledgments}
Support for H.A.S. and C.A.K. is provided by NSF grants AST 0607249 and AST 0707756.  Support for M.C. is provided by the Chilean Ministry for the Economy, Development, and Tourism's Programa Iniciativa Cient\'{i}fica Milenio through grant P07-021-F, awarded to The Milky Way Millennium Nucleus; by the BASAL Center for Astrophysics and Associated Technologies (PFB-06); by Proyecto Fondecyt Regular \#1110326; and by Proyecto Anillo ACT-86.

\end{document}